\documentclass[journal,comsoc]{IEEEtran}
\usepackage{graphicx}
\usepackage{subfigure}
\usepackage{float}
\usepackage{subfloat}
\usepackage{amsmath}
\usepackage{amssymb}
\usepackage{amsthm}
\usepackage{epstopdf}
\usepackage{cases}
\usepackage{color}
\usepackage{makecell}
\usepackage{cite}
\usepackage{booktabs}
\usepackage{diagbox}
\usepackage{multirow}
\usepackage{stfloats}
\usepackage{algorithmic}
\usepackage[ruled]{algorithm2e}
\usepackage{cases}
\usepackage{xcolor}

\usepackage[normalem]{ulem}

\usepackage{caption}
\usepackage{mathtools}
\usepackage[T1]{fontenc}
\ifCLASSINFOpdf

\else

\fi
\usepackage{amsmath}
\begin{document}

\title{Image Super-Resolution-Based Signal Enhancement in Bistatic ISAC}

\author{Yi~Wang,~\IEEEmembership{Member,~IEEE,}
        Keke~Zu,~\IEEEmembership{Member,~IEEE,}
        Luping~Xiang,~\IEEEmembership{Senior~Member,~IEEE,}
        Martin~Haardt,~\IEEEmembership{Fellow,~IEEE,}
        Chaochao~Wang,
        Xianchao~Zhang,~\IEEEmembership{Member,~IEEE,}
        and~Kun~Yang,~\IEEEmembership{Fellow,~IEEE}
\thanks{This work was supported in part by the Municipal Government of Quzhou under Grant (No. 2023D027, 2023D045, 2023D005), and in part by the Natural Science Foundation of China under Grant No. 62132004. (\emph{Corresponding author: Yi~Wang, Xianchao~Zhang.})}
\thanks{Y. Wang and K. Zu are with the Yangtze Delta Region Institute (Quzhou), University of Electronic Science and Technology of China, Quzhou 324003, China, e-mail: wangyi@csj.uestc.edu.cn, zukeke@csj.uestc.edu.cn.}
\thanks{L. Xiang is with the State Key Laboratory of Novel Software Technology, Nanjing University, Nanjing 210008, China, and also with the School of Intelligent Software, Engineering, Nanjing University (Suzhou Campus), Suzhou 215163, China, e-mail: luping.xiang@nju.edu.cn.}
\thanks{M. Haardt is with the Communications Research Laboratory, Ilmenau University of Technology, Ilmenau, Germany, email: martin.haardt@tu-ilmenau.de.}
\thanks{C. Wang and X. Zhang are with the Provincial Key Laboratory of Multimodal Perceiving and Intelligent Systems, Jiaxing University, Jiaxing 314000, China, email: ccwang@zjxu.edu.cn, zhangxianchao@zjxu.edu.cn.}
\thanks{K. Yang is with the State Key Laboratory of Novel Software Technology, Nanjing University, Nanjing, 210008, China, and School of Intelligent Software and Engineering, Nanjing University (Suzhou Campus), Suzhou, 215163, China, and School of Information and Communication Engineering, University of Electronic Science and Technology of China, Chengdu 611731, China, e-mail: kunyang@essex.ac.uk.}
}

\maketitle

\begin{abstract}
Bistatic Integrated Sensing And Communication (ISAC) is poised to become a cornerstone technology in next-generation communication networks, such as Beyond 5G (B5G) and 6G, by enabling the concurrent execution of sensing and communication functions without requiring significant modifications to existing infrastructure. Despite its promising potential, a major challenge in bistatic cooperative sensing lies in the degradation of sensing accuracy, primarily caused by the inherently weak received signals resulting from high reflection losses in complex environments. Traditional methods have predominantly relied on adaptive filtering techniques to enhance the Signal-to-Noise Ratio (SNR) by dynamically adjusting the filter coefficients. However, these methods often struggle to adapt effectively to the increasingly complex and diverse network topologies. To address these challenges, we propose a novel Image Super-Resolution-based Signal Enhancement (ISR-SE) framework that significantly improves the recognition and recovery capabilities of ISAC signals. Specifically, we first perform a time-frequency analysis by applying the Short-Time Fourier Transform (STFT) to the received signals, generating spectrograms that capture the frequency, magnitude, and phase components. These components are then mapped into RGB images, where each channel represents one of the extracted features, enabling a more intuitive and informative visualization of the signal structure. To enhance these RGB images, we design an improved denoising network that combines the strengths of the UNet architecture and diffusion models. This hybrid architecture leverages UNet's multi-scale feature extraction and the generative capacity of diffusion models to perform effective image denoising, thereby improving the quality and clarity of signal representations under low-SNR conditions. Following the image enhancement, ISAC signal processing is applied to the denoised outputs, enabling accurate extraction of transmitted communication symbols and precise estimation of target parameters such as range, velocity, and angle. Extensive simulation results demonstrate that our proposed ISR-SE method significantly improves the estimation accuracy by 63\% compared to traditional signal processing methods.


\end{abstract}


\IEEEpeerreviewmaketitle
\section{Introduction}
\IEEEPARstart{I}{n} many cities, Unmanned Aerial Vehicle (UAV) intrusion incidents continue to occur frequently, despite the potential for criminal prosecution for illegal UAV flights \cite{int1,int2,int3}. These unauthorized incursions pose significant safety risks to the public, particularly in critical areas such as airports, railways, and government facilities. Detecting UAVs with radar has several limitations, primarily due to the requirement for authorized spectrum, which requires prior applications and cannot ensure round-the-clock monitoring \cite{int4,int5}. In contrast, the Integrated Sensing And Communication (ISAC) technology, facilitated by 5G base stations, leverages existing 5G communication spectra, eliminating the need for additional spectrum \cite{int6,int7,int8}. This approach offers a high reliability and stability for UAV detection, enabling continuous, all weather, day-and-night surveillance.


Bistatic ISAC leverages existing communication infrastructures while minimizing the need for extensive hardware modifications, reducing deployment costs and accelerating the integration of advanced sensing capabilities into current systems \cite{int9,int10,int11}. Additionally, by strategically positioning transmitters and receivers, operators can optimize coverage areas and adapt to varying operational conditions. This adaptability is particularly beneficial in urban environments, where obstacles and signal reflections may hinder traditional radar systems. However, bistatic ISAC signals, which are often reflected from surrounding targets, encounter challenges such as noise interference, multipath fading, and other environmental disturbances \cite{int12,int13}. These issues can adversely affect signal clarity, resulting in lower detection and decoding accuracy. Moreover, with a small Radar Cross-Section (RCS) and reflection coefficient, this will further reduce the detectability of the intrusion UAVs \cite{int14,int15}. However, these conventional techniques struggle to cope with nonlinear and non-stationary noise, particularly under complex scenarios with weak signals, severely limiting the accuracy and reliability of target detection. Hence, there is an urgent need for a more robust and generalizable signal enhancement method to address these inherent shortcomings.

The common method to solve this problem is to use various filtering techniques to improve the quality and clarity of low-SNR signals \cite{AF1,AF2}. Adaptive filters dynamically adjust their parameters in real-time to optimize signal recovery, rendering them particularly suitable for environments where the noise characteristics are unpredictable or non-stationary. For example, the Least Mean Squares (LMS) algorithm updates the filter coefficients iteratively to minimize the mean squared error between the desired signal and the filter output \cite{LMS1,LMS2}. It is computationally efficient and widely used in applications like noise cancellation and echo suppression. The Recursive Least Squares (RLS) algorithm aims to minimize the sum of the squares of the error terms by recursively computing the filter coefficients \cite{RLS1,RLS2}. While it typically offers faster convergence and better performance in tracking signal variations compared to LMS, it requires more computational resources. However, as wireless environments become more complex and diverse, these conventional filtering techniques may exhibit limitations, particularly when dealing with non-stationary, nonlinear, or intricate background noise. The demand for Artificial Intelligence (AI) technology is greater than ever. Integrating AI into the ISAC system enhances signal processing efficiency, particularly in low-SNR scenarios.

By leveraging deep learning models, the system can effectively mitigate noise, extract essential features, and improve the accuracy of sensing and communication tasks. The study of \cite{CNN1} introduced a deep learning-based method for real-time noise removal in communication systems, employing Convolutional Neural Networks (CNNs) for single image super-resolution. Additionally, research demonstrated the use of deep convolutional neural networks to denoise scientific data, enabling the accurate detection of weak signals \cite{CNN2}. The study of \cite{CNN3} introduced a deep learning framework that integrates convolutional and recurrent neural networks to process time-series mobile sensing data, effectively handling sensor noise and enhancing feature extraction. Additionally, this capability not only addresses the limitations of traditional signal enhancement methods but also paves the way for intelligent networks. By leveraging these advancements, bistatic ISAC systems can achieve more reliable detection and interpretation of signals, ultimately enhancing their effectiveness across various applications.

With the development of AI, Generative Artificial Intelligence (GAI) excel in handling nonlinear, non-stationary, and high-noise scenarios, adapting to diverse signal environments and improving the detection and recognition of weak signals. The study of \cite{GAI1} designed Sea Clutter Suppression Generative Adversarial Network (SCS-GAN), which incorporates a Residual Attention Generator (RAG) and a Sea Clutter Discriminator (SCD) to suppress sea clutter in radar Plan-Position Indicator (PPI) images. This approach effectively reduces clutter while preserving marine target integrity. The study of \cite{GAI2} constructed a deep neural network comprising two Generative Adversarial Networks (GANs) to suppress clutter, thereby improving the signal-to-noise ratio while retaining target information. The study of \cite{GAI3} proposed Weak Target Enhancement Complex-valued GAN (WTE-CGAN) for target signal enhancement. This method introduces a target reconstruction generator and a target signal discriminator to extract complex-valued features of targets and clutter. By modifying the loss functions of the generator and discriminator, the generator is guided to more stably reconstruct target signals, significantly enhancing weak target signals. Pix2Pix is a type of Conditional GAN (CGAN) designed for image-to-image translation tasks, which learns a mapping from input images to output images using paired training data \cite{GAI4}. However, the adversarial nature of GANs, involving a generator and a discriminator in competition, often leads to unstable training dynamics. This instability can result in non-convergence, where the model fails to reach an equilibrium, making the training process unpredictable and challenging \cite{GAI5,GAI6}.

Although signal-to-image transformation and image enhancement using deep learning have been studied independently, this paper presents a novel task-driven framework that unifies them for the first time in the context of bistatic ISAC. We propose an RGB image encoding tailored for ISAC signal characteristics, and a hybrid UNet-diffusion denoising network adapted for complex noise and sparse low-SNR Radio Frequency (RF) signals. More importantly, we demonstrate that the generative priors learned in image space can significantly enhance the signal quality, outperforming traditional methods by a large margin. This work thus opens up a new paradigm of applying Generative Artificial Intelligence (GAI) for joint communication and sensing signal recovery in next-generation networks. The main contributions of this paper are summarized below:


\begin{itemize}
\item We propose a novel Image Super-Resolution-based Signal Enhancement (ISR-SE) framework that significantly improves the recognition and recovery capabilities of ISAC signals. Unlike traditional methods that rely on hand-crafted filters or statistical assumptions which often fail under extremely low-SNR conditions, the ISR-SE framework leverages deep generative models to learn hierarchical spectral representations from the time-frequency domain, which enables robust recognition and recovery under complex environmental conditions.

\item We propose an RGB image construction method and a diffusion-model-based image processing algorithm to enhance the signal quality in bistatic ISAC systems. Specifically, we transform the received signals into RGB images by mapping the critical features such as amplitude, frequency, and phase into distinct color channels. Subsequently, a diffusion model with an improved UNet denoising network architecture is employed to iteratively refine these images, effectively suppressing noise and enhancing feature representation, significantly improving the SNR of signal.

\item We propose an ISAC signal processing algorithm that integrates the Multiple Signal Classification (MUSIC) algorithm and the two-dimensional Discrete Fourier Transform (2D-DFT) passive sensing method to simultaneously extract angle, range, and velocity information. Specifically, the MUSIC algorithm is applied to accurately estimate the Angle of Arrival (AoA) by leveraging a spatial covariance analysis, while the 2D-DFT passive sensing algorithm extracts precise range and Doppler velocity information through frequency-domain processing.
\end{itemize}

\begin{figure}[!t]
	\centering
	\includegraphics[width=0.35\textheight]{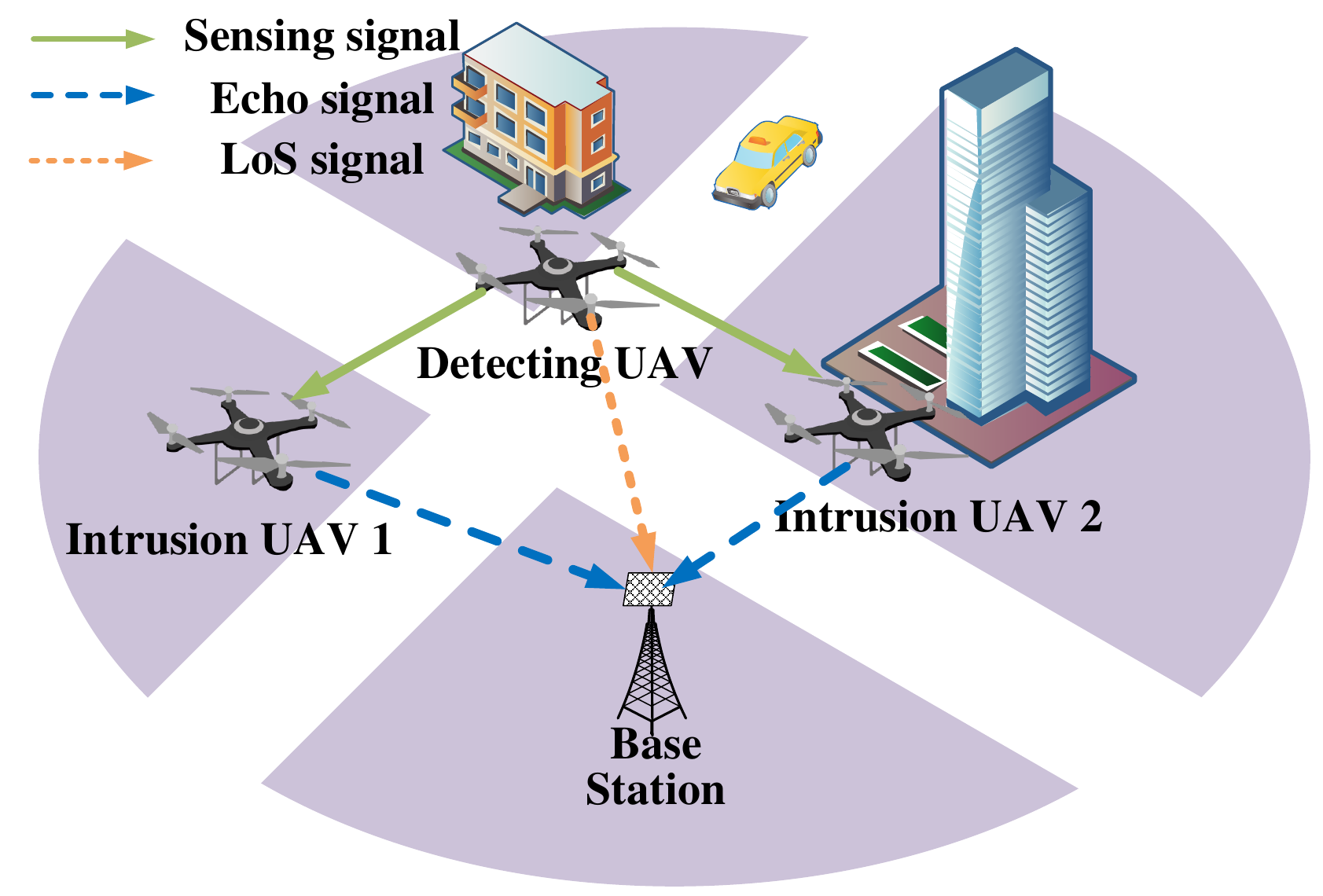}
	\DeclareGraphicsExtensions.
	\caption{The bistatic ISAC scenario.}
	\label{scen}
\end{figure}

The remainder of this article is structured as follows: Section II presents the system model of the bistatic ISAC system. Section III outlines the image super-resolution-based low-SNR signal enhancement method. Section IV details the signal processing of the ISAC signals. The estimation performance of the bistatic ISAC system is examined in Section V, with results provided. Finally, Section VI concludes the paper.

\textbf{Notations}: Bold uppercase letters denote matrices (e.g., $\boldsymbol{A}$). Bold lowercase letters denote column vectors (e.g., $\boldsymbol{v}$). Scalars are denoted by normal font (e.g., $x$); For an arbitrary matrix $\boldsymbol{A}$, rank($\boldsymbol{A}$), tr($\boldsymbol{A}$), $\boldsymbol{A^T}$, $\boldsymbol{A^H}$, and $\boldsymbol{A_{p,q}}$ denote its rank, trace, transpose, conjugate, and the $p$-th row and $q$-th column element, respectively. Moreover, ${\left\| {\cdot} \right\|}$ and ${\left| {\cdot} \right|}$ denote the Euclidean distance and the magnitude, respectively. The operator $ \otimes $, $ \odot $ and $ \circ $ denote the Kronecker product, Khatri-Rao product and Hadamard product, respectively.

\section{System Model}

\subsection{Scenario}
Consider a bistatic cooperative sensing scenario involving a ground BS and a detecting UAV, which together perform bistatic sensing operations, as illustrated in Fig.~\ref{scen}. The detecting UAV functions as an aerial base station, transmitting ISAC signals that include pilot signals and private data. These signals are reflected by nearby intrusion UAVs and subsequently received by the ground BS, enabling simultaneous data transmission and target detection. The communication signals are transmitted directly without containing any target sensing information, whereas the sensing signals, reflected by surrounding intrusion UAVs, are utilized for estimating the targets' distance, velocity, and angle through passive bistatic sensing.

We assume that both the BS and the detecting UAV remain physically stationary and that a dominant Line-of-Sight (LoS) path exists between them. Specifically, the distance between them and the AoA of the LoS path at the receiver are known, and these path parameters remain approximately stable within a short Coherent Processing Interval (CPI). Additionally, the BS and the detecting UAV operate in half-duplex mode. Furthermore, we assume that both the BS and the detecting UAV are equipped with two Uniform Planar Arrays (UPAs) for transmission and reception, respectively, with an inter-element spacing of half a wavelength. The three-dimensional position vectors of the BS and detecting UAV are represented as ${q_b} = {\left[ {{x_b},{y_b},{z_b}} \right]^T}$ and ${q_u} = {\left[ {{x_u},{y_u},{z_u}} \right]^T}$, respectively.

\subsection{Signal Model}
After propagating through the multipath channel, the ISAC signals transmitted from the BS consist of both LoS signals, which are directly used for communication, and NLoS signals, which are reflected by surrounding targets for simultaneous radar sensing. By utilizing the pilot signal for both communication channel estimation and passive bistatic sensing, the system effectively enables multi-functional use of the transmitted signals.

Since Orthogonal Frequency Division Multiplexing (OFDM) is widely adopted in current 5G systems and is expected to play a key role in future 6G systems, we employ OFDM as the transmitted signal model for the proposed ISAC system. The general baseband transmit signal is expressed as \cite{SM1}
\begin{equation}
s\left( t \right) = \sum\limits_{n = 0}^{N - 1} {\sum\limits_{m = 0}^{M - 1} {{s_{n,m}} \cdot {e^{j2\pi \left( {{f_c} + n\Delta f} \right)t}}\text{rect}\left( {\frac{{t - mT_s}}{T_s}} \right)} }
\end{equation}
where $M$ and $N$ denote the number of OFDM symbols and the number of subcarriers, respectively; ${s_{n,m}}$ denotes the transmit baseband symbol on the $n$-th subcarrier of the $m$-th OFDM symbol, ${{f_c} + {n\Delta f}}$ denotes the $n$-th subcarrier frequency and $\text{rect}\left( x \right)$ denotes the rectangle function. The subcarrier interval is ${\Delta f} $ and the OFDM symbol period is ${T_s} = \frac{1}{{\Delta f}} + {T_g}$ where ${T_g}$ is the period of cyclic prefix.

The transmit signal can be represented as \cite{SM2,SM2-1}
\begin{equation}
\boldsymbol{x \left( t \right)} = \boldsymbol{{w_t}}s\left( t \right),
\end{equation}
where $\boldsymbol{{w_t}}$ denotes the transmit beamforming vector. The beamforming vectors for communication and sensing can be generated by using a generalized Least Square (LS) method respectively according to the desired beam patterns. Then a multi-beam design for ISAC can be generated by combining the two beamforming vectors through a phase shifting term ${e^{j\varphi }}$ and a power distribution factor $0 \le {\beta _R} \le 1$. Hence, $\boldsymbol{w_t }$ can be written as
\begin{equation}
\boldsymbol{{w_t}} = \sqrt {{\beta _R}} {e^{j\varphi }}{\boldsymbol{w_{t,s}}} + \sqrt {1 - {\beta _R}} {\boldsymbol{w_{t,c}}},
\end{equation}
where $\boldsymbol{{w_{t,s}}}$, $\left\| {\boldsymbol{{w_{t,s}}}} \right\|_2^2 = 1$, and $\boldsymbol{{w_{t,c}}}$, $\left\| {\boldsymbol{{w_{t,c}}}} \right\|_2^2 = 1$ denote two beamforming vectors of the scanning and fixed sub-beams, respectively.

Since UAVs typically operate at relatively high altitudes, the communication links between the detecting UAV and the BS are primarily dominated by the LoS component. Therefore, assuming the presence of a dominant LoS path between them is reasonable. Based on the Rician channel model for UAV-enabled ISAC networks \cite{SM3}, it can be expressed as
\begin{equation}
\boldsymbol{{H}} = \frac{{\sqrt {{K_R}} }}{{\sqrt {{K_R} + 1} }}\boldsymbol{H^{LoS}} + \frac{1}{{\sqrt {{K_R} + 1} }}\boldsymbol{H^{NLoS}},
\end{equation}
where
\begin{equation}
{\boldsymbol{H^{LoS}} = {\beta _{0}}{e^{j2\pi {f_{0}}t}}\delta \left( {t - {\tau _{0}}} \right)\boldsymbol{\alpha \left( {{q}_{r,0}} \right){\alpha ^T}\left( {{q}_{t,0}} \right)}}
\end{equation}
and
\begin{equation}
{\boldsymbol{H^{NLoS}} = \sum\limits_{l = 1}^L {{\beta _{l}}{e^{j2\pi {f_{l}}t}}\delta \left( {t - {\tau _{l}}} \right)\boldsymbol{\alpha \left( {{q}_{r,l}} \right){\alpha ^T}\left( {{q}_{t,l}} \right)} }}
\end{equation}
denote the LoS component and the NLoS components of the channel, respectively. Moreover, ${K_R}$ denotes the Rician K-factor, ${L}$ denotes the number of multipaths, and ${{\beta _{l}}}$, ${{f_{l}}}$, ${{\tau _{l}}}$ denote the channel fading coefficient, Doppler frequency, and delay of the $l \in \left\{ {0,1,2, \cdots ,{L}} \right\}$-th path, respectively; $\boldsymbol{{{q}_{r,l}}}$ and $\boldsymbol{{{q}_{t,l}}}$ are the AoA and Angle of Departure (AoD), respectively. Furthermore, ${\beta _{0}} = \sqrt {\frac{{{\lambda ^2}}}{{{{\left( {4\pi {d_{L}}} \right)}^2}}}}$ and ${\beta _{l}} = \sqrt {\frac{{{\lambda ^2}}}{{{{\left( {4\pi } \right)}^3}d_{NL,1}^2d_{NL,2}^2}}} {\rho _{l}}$ denote the attenuations of the LoS path and the NLoS paths, respectively; ${\rho _{l}}$ denotes the reflecting factor of the $l$-th path, with $d_{NL,1}$ and $d_{NL,2}$ being the corresponding distances. For an array of $M_x^b \times M_y^b$ antennas with an AoA/AoD $\boldsymbol{{{{q}}_k}} = {\left( {{\varphi _k},{\theta _k}} \right)^T}$, the array steering vector can be expressed as
\begin{equation}
\boldsymbol{\alpha\left( {{{{q}}_k}} \right)} = \left[ {\begin{array}{*{20}{c}}
1\\
{{e^{-j\pi \Omega _y^k}}}\\
 \vdots \\
{{e^{-j\left( {M_y^b - 1} \right)\pi \Omega _y^k}}}
\end{array}} \right] \otimes \left[ {\begin{array}{*{20}{c}}
1\\
{{e^{-j\pi \Omega _x^k}}}\\
 \vdots \\
{{e^{-j\left( {M_x^b - 1} \right)\pi \Omega _x^k}}}
\end{array}} \right],
\end{equation}
where $\Omega _y^k = \sin {\theta _k}\sin {\varphi _k}$, $\Omega _x^k = \sin {\theta _k}\cos {\varphi _k}$, the operator $\otimes$ denotes the Kronecker product, and ${\varphi _k}$ and ${\theta _k}$ are the azimuth angle and elevation angle, respectively.

The received signal for either sensing or communication is thus given by
\begin{equation}
\begin{array}{l}
\begin{aligned}
&\boldsymbol{{y_{n,m}}} = \boldsymbol{{H_{n,m}}{w_t}}{s_{n,m}} + \boldsymbol{{z_{n,m}}}\\
&= \sum\limits_{l = 0}^L {{\beta _l}{\chi _{t,l}}{e^{ - j2\pi n\Delta f{\tau _l}}} \cdot {e^{j2\pi {f_l}m{T_s}}} \cdot } \boldsymbol{\alpha \left( {{q_{r,l}}} \right)}{s_{n,m}} + \boldsymbol{{z_{n,m}}}
\end{aligned}
\end{array},
\end{equation}
where $\boldsymbol{{H_{n,m}}}$ denotes the channel response on the $n$-th subcarrier of the $m$-th packet, ${\chi _{t,l}} = \boldsymbol{{\alpha ^T}\left( {{q_{t,l}}} \right){w_t}}$ denotes the the transmit beamforming gain, and $\boldsymbol{{z_{n,m}}}$ denotes the Additive White Gaussion Noise (AWGN).



As for the communication, the directly conveyed communication path does not convey any sensing information. As for the sensing, after propagation through the multipath channel, the motion state of targets can be estimated using bi-static sensing. Since the signal strength after multiple reflections is negligible, we only consider echoes that are directly reflected from the targets. However, due to high reflection loss, the reflected signals are typically weak, leading to a decrease in sensing performance, which in turn affects the accuracy of targets' parameters estimation.


\begin{figure*}[t!]
	\centering
	\includegraphics[width=0.55\textwidth]{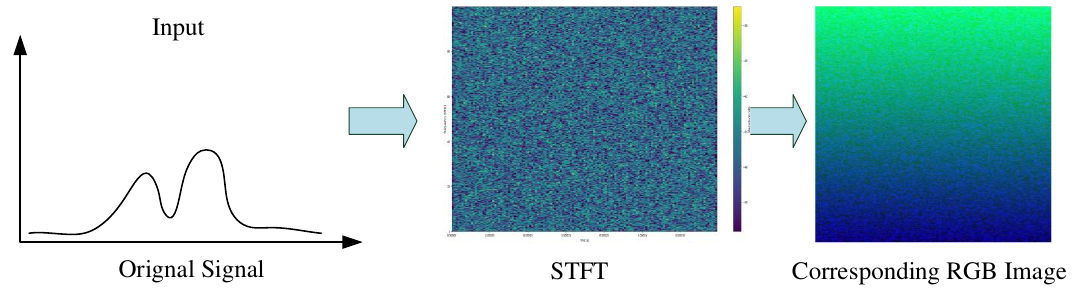} %
	\caption{The construction of RGB Image.}
	\label{const}
\end{figure*}

\section{Image Super-Resolution-Based Signal Enhancement Framework}
In bistatic sensing, the echo signals are first transmitted and then reflected from surrounding targets before being received by a separate receiver. Due to the free space propagation loss and the low reflectivity of certain targets, the received signals often exhibit significantly reduced power, leading to a low Signal-to-Noise Ratio (SNR). This degradation in signal quality can severely impact the sensing performance, making it challenging to accurately detect and localize targets. To mitigate the sensing performance degradation caused by low SNR conditions, we propose an Image Super-Resolution-based Signal Enhancement (ISR-SE) framework that leverages the generative capabilities of Generative Artificial Intelligence (GAI) models.
In our proposed method, the received low-SNR signals are first transformed into time-frequency representations, typically via STFT, to capture their spectral and temporal characteristics. These representations are then encoded into multi-channel image-like structures, such as RGB maps, where amplitude, frequency, and phase components are distinctly separated. The resulting images are subsequently processed by a diffusion model-based image processing network, which reconstructs fine-grained spectral details and suppresses background noise, thus significantly improving the quality of the signal for downstream sensing and communication tasks.

\subsection{RGB Image Construction}
In this subsection, we aim to convert low SNR complex signals into RGB images, after that employ the diffusion model to improve the quality of the RGB images, effectively boosting the power of the sensing signals.

After receiving the echo signals, we process them to generate Short-Time Fourier Transform (STFT) spectrograms, which are then used to construct RGB images that encode key spectral features, including frequency, magnitude, and phase information. Mapping frequency, amplitude, and phase into RGB channels ensures complete preservation of critical time-frequency information, thereby fully exploiting spatial correlations inherent in image-based processing methods. As shown in Fig.~\ref{const}, the process of converting the received signal into an RGB representation involves several critical steps:

1) Preprocessing and Windowing: The received signal is first segmented using a Hamming window function with a window length of 128 samples and an overlap of 120 samples. This windowing operation ensures smooth transitions between successive segments and helps mitigate spectral leakage when transforming the signal into the frequency domain. The overlapping segments provide better temporal resolution while maintaining sufficient frequency resolution.

2) Short-Time Fourier Transform (STFT): The STFT is computed for each windowed segment to obtain the time-frequency representation of the received signal. The STFT is defined as
\begin{equation}
Y\left( {t,f} \right) = \sum\limits_n {y\left( n \right){w_{ham}}\left( {n - t} \right){e^{ - j2\pi fn}}} ,
\end{equation}
where ${y\left( n \right)}$ denotes the input signal and ${{w_{ham}}\left( {n } \right)}$ denotes the Hamming window function. The output of the STFT consists of complex-valued coefficients at different time and frequency points, which can be decomposed into two primary components, i.e., the magnitude spectrum at each time-frequency bin and the phase spectrum.

3) Feature Extraction and Mapping: To construct an RGB image representation of the signal, we map the extracted spectral information into three separate color channels to form a comprehensive visual representation of the signal characteristics.

\emph{Red Channel (R)}: Encodes the magnitude spectrum, which is normalized to a range of [0,255] using logarithmic scaling to enhance the visibility of weaker signals.

\emph{Green Channel (G)}: Encodes the frequency component, typically by normalizing the frequency indices to [0,255] across the image dimensions.

\emph{Blue Channel (B)}: Encodes the phase spectrum, often by normalizing it to [0,255].

4) Image Construction: The three normalized matrices (frequency, magnitude, and phase) are then stacked along the channel dimension to form a three-channel RGB image. The resulting image provides an intuitive visualization of the received signal's spectral characteristics, making it suitable for subsequent deep learning or pattern recognition tasks.

Following this structured pipeline, we effectively transform the received echo signals into meaningful RGB representations that preserve crucial spectral information, facilitating further signal analysis and machine learning-based feature extraction. Since the spectral structure and modulation characteristics of a signal are transformed into image-like textures through a time-frequency transformation (e.g., STFT), the task of signal enhancement can be reframed as an image restoration problem.

\begin{figure*}[!t]
	\centering
	\includegraphics[width=0.6\textheight]{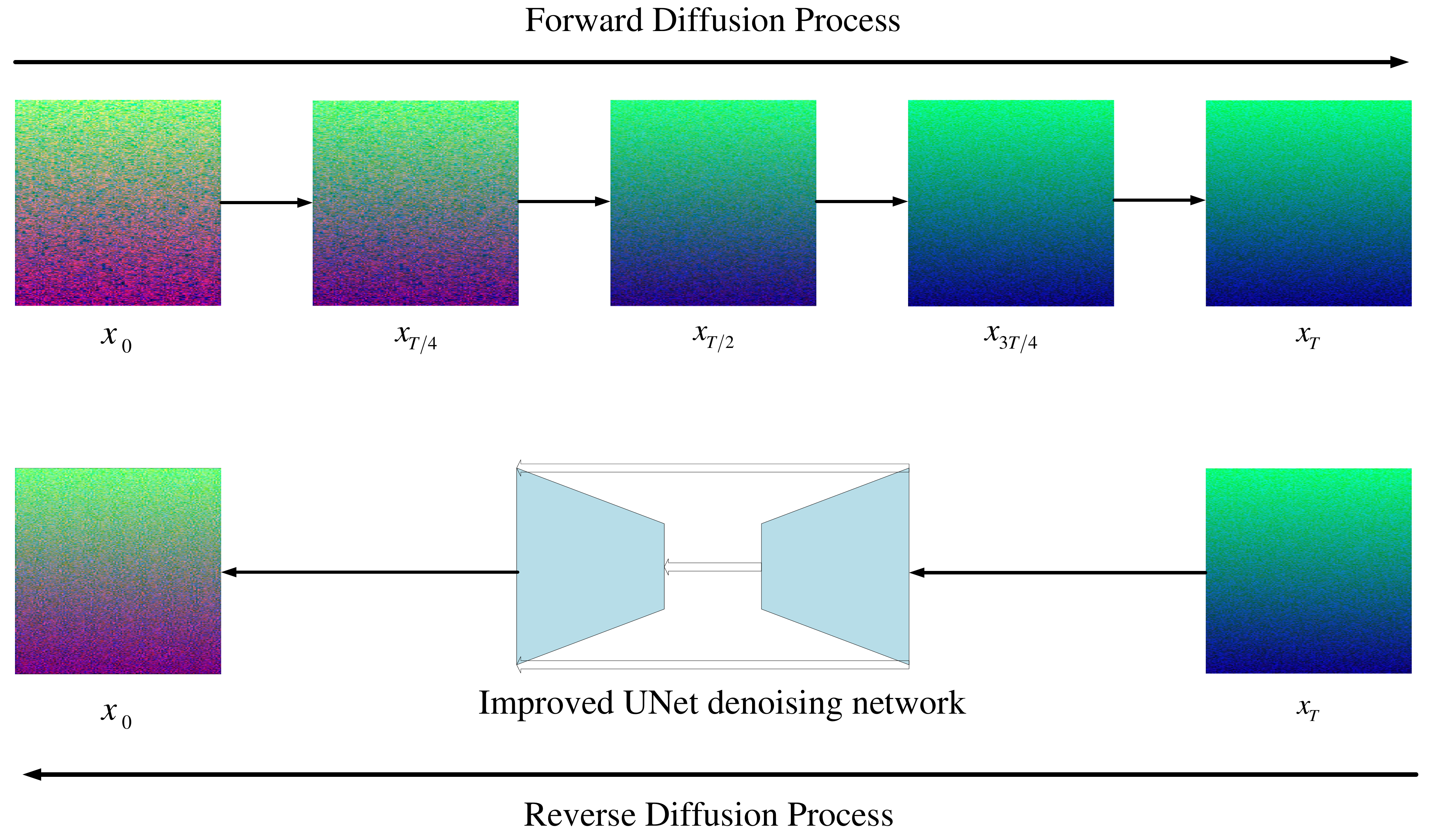}
	\DeclareGraphicsExtensions.
	\caption{The example of diffusion model.}
	\label{diff}
\end{figure*}

\subsection{Diffusion Model-Based RGB Image Processing}
With the rapid advancement of deep learning, generative AI models have achieved significant success in various image processing tasks. Among these, diffusion models-a newer class of generative AI models-have shown remarkable promise, particularly in image denoising, due to their excellent performance in the reverse diffusion process \cite{DM1,DM2,DM3}. In this section, we explore the application of diffusion models for ISR-SR framework, outlining their architecture and analyzing the underlying principles.


As shown in Fig.~\ref{diff}, this approach consists of two main steps: the forward diffusion process and the reverse diffusion process. In the forward diffusion process, the model progressively ``diffuses'' the image information, transforming it into noise. This typically involves the gradual addition of Gaussian noise until the image becomes random noise. In the reverse diffusion process, which is the inverse of the forward diffusion process, the model attempts to reconstruct high-quality image from the noise.

\emph{Forward Diffusion Process:} Assuming the original RGB image is $\boldsymbol{x_0}$, after $T$ forward steps, the original RGB image becomes $\boldsymbol{x_T}$, which approaches pure random noise. At each step $t$, the image $\boldsymbol{x_t}$ is generated by applying a controlled amount of Gaussian noise to the image from the previous step $\boldsymbol{x_{t-1}}$, following the equation:
\begin{equation}
q\left( {\boldsymbol{{x_t}|{x_{t - 1}}}} \right) = \mathcal{N}\left( {\boldsymbol{{x_t}};\boldsymbol{{\mu _t}} = \sqrt {1 - {\beta _t}} \boldsymbol{{x_{t - 1}}},\boldsymbol{\sum\nolimits_t }={{\beta _t}\boldsymbol{\rm I}} } \right),
\end{equation}
where $q\left( {\boldsymbol{{x_t}|{x_{t - 1}}}} \right)$ denotes a normal distribution, characterized by the mean $\boldsymbol{{\mu _t}}$ and the variance $\boldsymbol{\sum\nolimits_t }$, $\boldsymbol{\rm I}$ denotes the identity matrix, and $\beta _t$ denotes a predefined noise variance schedule that determines how much noise is added at each step. This can be seen as a Markov process, where each step involves adding a certain degree of noise to the current image. The number of corruption steps, denoted as $T$, determines how gradually the image transitions from its original form to pure noise.

\emph{Reverse Diffusion Process:} A deep neural network (e.g., a UNet) is trained to iteratively predict and remove noise at each step, progressively reconstructing the noisy image until it closely resembles the original clean image. This process is a reverse Markov process, following the equation:
\begin{equation}
{p_\theta }\left( \boldsymbol{{{x_{t - 1}}|{x_t}}} \right) = \mathcal{N}\left( {\boldsymbol{{x_{t - 1}}};\boldsymbol{{\mu _\theta }}\left( {\boldsymbol{{x_t}},t} \right),\boldsymbol{\sum\nolimits_\theta } {\left( {\boldsymbol{{x_t}},t} \right)} } \right),
\end{equation}
where
\begin{equation}
\boldsymbol{{\mu _\theta }}\left( {\boldsymbol{{x_t}},t} \right) = \frac{1}{{\sqrt {{\alpha _t}} }}\left( {\boldsymbol{{x_t}} - \frac{{{\beta _t}}}{{\sqrt {1 - {\overline{\alpha} _t}} }}\boldsymbol{{\hat z_{f,\theta}}}\left( {\boldsymbol{{x_t}},t} \right)} \right),
\end{equation}
${\alpha _t} = 1- {\beta _t}$, ${\overline{\alpha} _t} = \prod\limits_{i = 1}^t {{\alpha _i}} $,$\boldsymbol{{\hat z_{f,\theta}}}\left( {\boldsymbol{{x_t}},t} \right)$ denotes the predicted noise.
By conditioning the model on time step $t$, it can learn to predict the Gaussian parameters, i.e., the mean $\boldsymbol{{\mu _\theta }}\left( {\boldsymbol{{x_t}},t} \right)$ and the covariance matrix $\boldsymbol{\sum\nolimits_\theta } {\left( {\boldsymbol{{x_t}},t} \right)} $ for each time step. The complete reverse diffusion process is shown in Algorithm~\ref{RDP}.

\begin{algorithm}[!ht]
	\caption{Reverse Diffusion Process} \label{RDP}
	\begin{algorithmic}[1]
\STATE $\boldsymbol{x_T}  \in \mathcal{N}\left( {\boldsymbol{0,\rm I}} \right)$
\FOR {$t = T, \cdots ,1$}
\STATE $\boldsymbol{z}  \in \mathcal{N}\left( {\boldsymbol{0,\rm I}} \right)$ if $t > 1$, else $\boldsymbol{z =0}$
\STATE $\boldsymbol{{x_{t - 1}}} = \frac{1}{{\sqrt {{\alpha _t}} }}\left( {\boldsymbol{{x_t}} - \frac{{{\beta _t}}}{{\sqrt {1 - {\overline{\alpha} _t}} }}\boldsymbol{{\hat z_f}}\left( {\boldsymbol{{x_t}},t} \right)} \right) + {\sigma _t}\boldsymbol{z}$
\ENDFOR
\STATE \textbf{return $\boldsymbol{x_0}$}
	\end{algorithmic}
\end{algorithm}


\begin{figure}[!t]
	\centering
	\includegraphics[width=0.35\textheight]{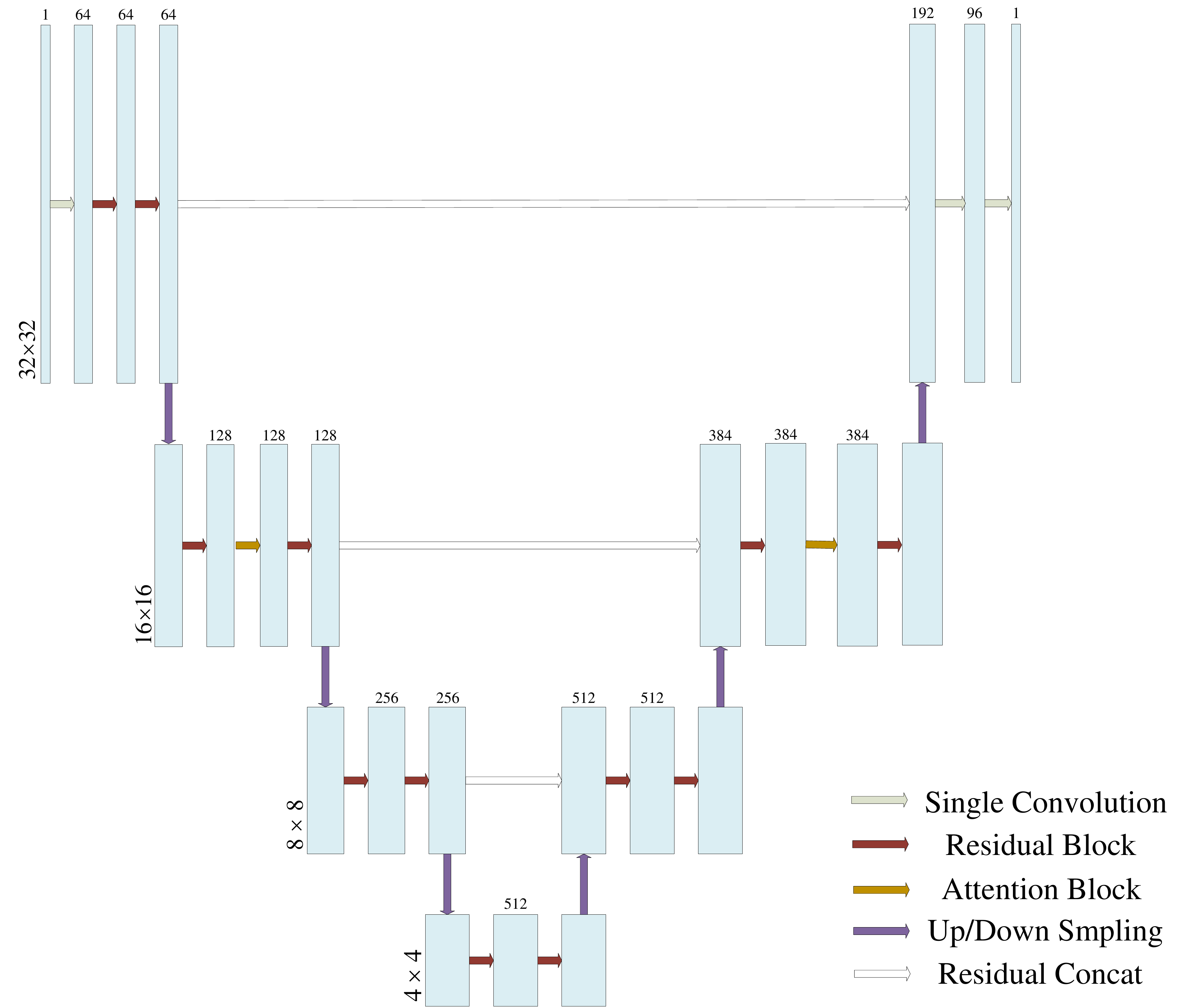}
	\DeclareGraphicsExtensions.
	\caption{The improved UNet denoising network architecture.}
	\label{denoise}
\end{figure}

The key challenge is to design an accurate denoising network architecture to estimate the noise component in the corrupted image and refines the image step by step. The UNet is a classic image segmentation network, characterized by its encoder-decoder structure. The encoder progressively reduces the spatial resolution of the image while increasing the number of channels, thereby capturing deep hierarchical features. The decoder performs the opposite operation, gradually restoring the spatial resolution while decreasing the number of channels, ultimately producing an output image of the same size as the input.
Despite its success in various image restoration tasks, the conventional UNet architecture exhibits several limitations when applied in ISAC scenarios. First, although the UNet utilizes skip connections to fuse high-resolution shallow features with deeper semantic representations, these connections become insufficient when the network depth increases. Moreover, the original UNet treats all features equally across spatial and channel dimensions, lacking the ability to automatically focus on the most informative components of the signal.

As shown in Fig.~\ref{denoise}, we propose an improved UNet denoising network architecture to conduct image denoising tasks that leverages the strengths UNet, residual network, and self-attention mechanisms, which not only incorporates the traditional UNet structure but also integrates residual blocks and an attention mechanism to enhance noise prediction at each step of the image denoising process. By employing a combination of residual connection and feature concatenation, Residual Cancat not only preserves high-resolution features extracted by the encoder and passes them directly to the decoder, but also addresses the vanishing problem in deep networks. Additionally, a channel attention mechanism is employed to dynamically adjust the importance of each channel, enabling the network to focus on the most relevant regions for the denoising task. By automatically emphasizing critical features, the model enhances its ability to remove noise while preserving essential image details.

\begin{figure*}[t!]
	\centering
	\includegraphics[width=0.55\textwidth]{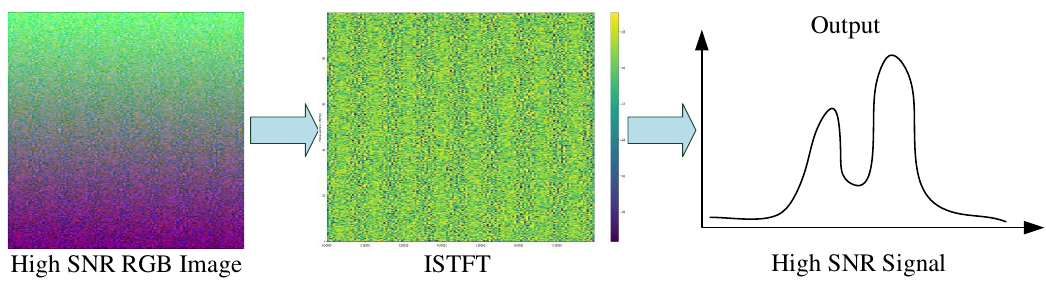} %
	\caption{The reconstruction of high SNR signal.}
	\label{rever}
\end{figure*}


The training objective is to teach the neural network, implemented using a UNet architecture enhanced with residual network and attention mechanisms, to accurately estimate and subsequently remove the noise introduced at each incremental step of the diffusion process. Specifically, the training objective is designed as a composite loss function incorporating both the Mean Squared Error (MSE) of estimated the noise and the Structural Similarity Index (SSIM). Specifically, the objective function can be mathematically formulated as
\begin{equation}
\mathcal{L} = \alpha  \cdot \rm MSE\left( {\boldsymbol{{ z_{f,\theta}}},\boldsymbol{{\hat z_{f,\theta}}}} \right) + \left( {1 - \alpha } \right) \cdot \left( {1 - \rm SSIM\left( {\boldsymbol{{x_t}},{\boldsymbol{{\hat x}_t}}} \right)} \right).
\end{equation}
The MSE quantifies the pixel-wise difference between the original and the estimated noise, which is computed as
\begin{equation}
\rm MSE\left( {\boldsymbol{{z_{f,\theta }}},{\boldsymbol{{\hat z}_{f,\theta }}}} \right) = \frac{1}{\mathit{N}}\sum\limits_{\mathit{i} = 1}^\mathit{N} {{{\left( {\boldsymbol{{z_{f,\theta,i }}} - {\boldsymbol{{\hat z}_{f,\theta,i }}}} \right)}^2}},
\end{equation}
where $N$ denotes the total number of sampled estimated noise values.
The SSIM measures the perceptual similarity between two images based on structural information, luminance, and contrast, which is computed as
\begin{equation}
\rm SSIM\left(\boldsymbol{x},\boldsymbol{\hat x}\right) = \frac{{\left(2{\mu _{\boldsymbol{x}}}{\mu _{\boldsymbol{\hat x}}} + {c_1}\right)\left(2{\sigma _{\boldsymbol{x\hat x}}} + {c_2}\right)}}{{\left(\mu _{\boldsymbol{x}}^2 + \mu _{\boldsymbol{\hat x}}^2 + {c_1}\right)\left(\sigma _{\boldsymbol{x}}^2 + \sigma _{\boldsymbol{\hat x}}^2 + {c_2}\right)}},
\end{equation}
where ${\mu _{\boldsymbol{x}}}$ and ${\mu _{\boldsymbol{\hat x}}}$ denote the average brightness of the original image ${\boldsymbol{x}}$ and the estimated image $\boldsymbol{\hat x}$, $\sigma _{\boldsymbol{x}}^2$ and $\sigma _{\boldsymbol{\hat x}}^2$ denote the variance of the two images, ${\sigma _{\boldsymbol{x\hat x}}}$ denotes the cross-covariance of the two images, and ${c_1}$ and ${c_2}$ constants to avoid situations where the denominator is zero.
The hyperparameter $\alpha  \in \left[ {0,1} \right]$ balances the contributions of the pixel-wise accuracy (captured by the MSE) and the structural perceptual quality (captured by the SSIM).

Through this iterative training procedure, the neural network gains the capability to reverse the diffusion steps, reconstructing high-quality, clean images from their heavily corrupted, noisy counterparts. Each step requires a full forward pass through a deep generative model, commonly based on a modified UNet architecture. The computational complexity per step can be approximated as $\mathcal{O}(dHW)$, where $d$, $H$, $W$ denote the channel number, image height, and width, respectively.

As shown in Fig.~\ref{rever}, after reconstructing the high SNR image, the resulting pixel values from the RGB channels represent the enhanced magnitude, frequency, and phase information obtained during the diffusion-based denoising procedure. These RGB values are then mapped back to their corresponding complex spectral representations. Specifically, by converting the RGB channels into amplitude, frequency, and phase components, we reconstruct the complete time-frequency representation of the denoised signal. Subsequently, we apply the Inverse Short-Time Fourier Transform (ISTFT) to this reconstructed spectrogram, enabling the recovery of the enhanced time-domain signal. The ISTFT spectrogram accurately reflects the improved signal quality, capturing the fine-grained structural details and effectively suppressing noise interference. Consequently, the entire process yields a significantly enhanced output signal characterized by a high signal-to-noise ratio, improved perceptual quality, and accurate preservation of essential temporal and spectral features.

Diffusion model-based image super-resolution is particularly effective at recovering fine-grained textures and structural patterns in the image domain, which directly correspond to signal features such as amplitude, frequency content, and phase continuity. Processing in the image domain therefore translates to frequency-domain enhancement and noise suppression in the signal space. Furthermore, such models enable the incorporation of complex statistical priors and nonlinear representations, which go far beyond the capabilities of traditional linear filters, allowing for robust denoising even under extremely low-SNR conditions.

\subsection{Summary}
\begin{algorithm}[!ht]
\caption{ISR-based Signal Enhancement Framewrok}\label{ISRA}
\begin{algorithmic}[1]
\STATE \textbf{Input}: Received signal $\boldsymbol{{y_{n,m}}}$, pretrained diffusion model $G$.
\STATE \textbf{Output}: Enhanced signal $\boldsymbol{{\hat{y}_{n,m}}}$.
\STATE \textbf{Step 1: STFT}
\STATE Perform STFT on $\boldsymbol{{y_{n,m}}}$, and extract the magnitude, frequency, and phased information.
\STATE \textbf{Step 2: RGB image construction}
\STATE Map the extracted spectral information into three separate color channels after normalization. Construct RGB image $\boldsymbol{x_T}$.
\STATE \textbf{Step 3: Diffusion model-based image processing}
\STATE Feed $\boldsymbol{x_T}$ into pretrained diffusion model $G$, and perform the image super-resolution operation according to Algorithm~\ref{RDP}. Recover super-resolution image $\boldsymbol{x_0}$.
\STATE \textbf{Step 4: High SNR signal reconstruction}
\STATE Perform ISTFT to reconstruct time-domain signal $\boldsymbol{{\hat{y}_{n,m}}}$.
\STATE \textbf{return $\boldsymbol{{\hat{y}_{n,m}}}$.}
\end{algorithmic}
\end{algorithm}

To address the degradation of the sensing and communication performance under low SNR conditions, we propose an ISR-SE framework, which leverages the generative capabilities of deep learning models, especially diffusion model-based architectures. As shown in Algorithm~\ref{ISRA}, the proposed method consists of the following major steps:

1) STFT: The received low-SNR signal $\boldsymbol{{y_{n,m}}}$ is first transformed into the time-frequency domain using the STFT. This process generates a complex spectrogram, from which we extract three critical components: the magnitude spectrum, frequency, and phase information. These components encapsulate the primary features used for both signal interpretation and image construction.

2) RGB Image Construction: The extracted spectral features are normalized and mapped into a three-channel RGB image.
\begin{itemize}
\item The Red (R) channel represents the normalized logarithmic magnitude spectrum.
    \begin{equation}
    R = \frac{{\log \left( {\left| {Y\left( {f,t} \right)} \right| + \varepsilon } \right)}}{{\log \left( {\left| {{M_{\max }}} \right| + \varepsilon } \right)}} \times 255,
    \end{equation}
    where ${M_{\max }}$ denotes the expected maximum magnitude observed at high SNR and $\varepsilon$ is a small value to avoid ${\log \left( 0 \right)}$.
\item The Green (G) channel encodes the normalized frequency index.
\begin{equation}
G = \frac{{f - {f_{\min }}}}{{{f_{max}} - {f_{\min }}}} \times 255,
\end{equation}
where the choice of ${f_{\min }}$ and ${f_{\max }}$ is determined by the sampling frequency and STFT configuration.
\item The Blue (B) channel carries the normalized phase information.
\begin{equation}
B = \frac{{\angle Y\left( {f,t} \right) + \pi }}{{2\pi }} \times 255.
\end{equation}
The Phase is naturally bounded between $\left[ { - \pi ,\pi } \right]$, so the transformation maps it to $\left[ {0,255} \right]$.
\end{itemize}

This image $\boldsymbol{x_T}$ serves as a compact and structured representation of the original signal in a visual domain that is well-suited for deep learning models.

3) Diffusion model-based image processing: The constructed RGB image is then fed into a diffusion model-based super-resolution network $G$. This model learns to suppress noise and reconstruct missing spectral details by capturing multi-scale spatial dependencies and incorporating strong statistical priors. The output is an enhanced image $\boldsymbol{x_0}$, which visually restores signal texture that correlates with key signal characteristics.

4) High SNR signal reconstruction: The ISTFT is applied to recover the high-SNR time-domain signal $\boldsymbol{{\hat{y}_{n,m}}}$, which exhibits improved fidelity, reduced noise, and enhanced resolution for both sensing and communication tasks.

Through this image-domain processing pipeline, the proposed ISR-SE framework transforms the signal enhancement problem into a visual inference task, enabling the use of powerful image super-resolution techniques to restore weak signals.

\section{ISAC Signal Processing}
After obtaining the high-SNR time-domain signal, comprehensive utilization of spatial, temporal, and frequency-domain information can effectively support both the demodulation of the transmitted communication signals and precise estimation of the parameters of the targets (i.e., angle, range and velocity). Specifically, the received signal undergoes demodulation to recover the embedded communication data, ensuring reliable information exchange. Simultaneously, the channel estimate on the $n$-th subcarrier of the $m$-th packet can be expressed as
\begin{equation}\label{esta}
\begin{array}{l}
\begin{aligned}
&\boldsymbol{{\hat{h}_{n,m}}} = \boldsymbol{{h_{n,m}}} + \boldsymbol{z_{n,m,h}}\\
 &= \sum\limits_{l = 0}^L {{\beta _{l}}{e^{j2\pi m{T_s} {{f_{l}^e} } }} \times } {e^{ - j2\pi n\Delta f {{\tau _{l}^e}} }}   {\chi _{t,k}}\boldsymbol{{\alpha}\left( {q_{r,l}} \right)}
 + \boldsymbol{z_{n,m,h}} ,
 \end{aligned}
\end{array}
\end{equation}
where $\boldsymbol{{h_{n,m}}} \in {\mathbb{C}^{{M_x^b  M_y^b }  \times  1 }}$, ${\chi _{t,k}} = \boldsymbol{{\alpha^T}\left( {q_{t,l}} \right){w_t}}$ denotes the transmit beamforming gain, ${f_{l}^e }$ and ${\tau _{l}^e}$ denote the Doppler frequency and the delay of $l$-th path, respectively; and $\boldsymbol{z_{n,m,h}}$ denotes the Gaussian noise vector with variance $\sigma _h^2$. For simplicity, we assume that only one channel estimation is performed on each packet and $M_s$ CSI estimates are obtained for bistatic sensing during an interval of $T_p^s$ where $M_s$ denotes the number of OFDM packets. We also assume that there are ${P_s}$ OFDM symbols in each packet and the interval of CSI estimation is $T_p^s = {P_s}{T_s}$.


The estimation of the AoAs is performed by employing the spatial smoothing MUSIC algorithm. This method partitions a UPA into multiple overlapping subarrays and computes the covariance matrix for each subarray individually. By averaging these covariance matrices, spatial smoothing effectively restores the rank of the overall covariance matrix, thereby decorrelating the incident signals. The resulting smoothed covariance matrix can then be used in the standard MUSIC framework to accurately estimate the AoAs. By stacking all $M_s \times N$ channel estimates, the estimated correlation matrix of $\boldsymbol{{\boldsymbol{\hat{H}}}} \in {\mathbb{C}^{{M_x^b  M_y^b }  \times  {M_s}N }}$ is written as
\begin{equation}
\boldsymbol{{R_{h}}} = {{{\boldsymbol{\hat{H}}}\boldsymbol{\hat{H}^H}} \mathord{\left/
 {\vphantom {h {\left( {M \times N} \right)}}} \right.
 \kern-\nulldelimiterspace} {\left( {M_s N} \right)}}  ,
\end{equation}
where
\begin{equation}
{{\boldsymbol{\hat{H}}}\boldsymbol{\hat{H}^H}} = \sum\limits_{n = 1}^N {\sum\limits_{m = 1}^{{M_s}} {{\boldsymbol{{\hat{h}_{n,m}}}\boldsymbol{{\hat{h}_{n,m}^H}}} }  } .
\end{equation}
After spatial smoothing, the AoAs can be estimated by identifying the peaks of the MUSIC spatial spectrum. Additionally, the estimation of range and velocity are obtained through the 2D-DFT sensing algorithm, which can achieve a relatively high sensing performance with a low complexity. It is assumed that the clock offset between the transmitter and receiver has been corrected, and accurate time and frequency synchronization have been realized. To steer the receive beams toward the dominant signal paths and effectively separate the multipath components, we adopt a Least Squares (LS) approach to design beamforming vectors corresponding to each estimated AoA at the BS. The desired array response matrix is constructed over $K_s$ target directions, where $K_s$ is determined based on the number of AoAs resolved by the MUSIC algorithm. Specifically, let $\boldsymbol{A_q} = \left[ {\boldsymbol{\alpha \left( {{q_1}} \right)},\boldsymbol{\alpha \left( {{q_2}} \right)}, \cdots ,\boldsymbol{\alpha \left( {{q_{{K_s}}}} \right)}} \right]^T \in \mathbb{C}^{K_s \times M_x^b M_y^b }$ denote the actual array response matrix at the BS, where each row corresponds to the steering vector associated with a specific spatial direction among the $K_s$ estimated AoAs.
Let $\boldsymbol{v} = \left[ {{v_1},{v_2}, \cdots ,{v_{{K_s}}}} \right] ^T \in \mathbb{C}^{K_s \times 1} $ represent the desired array response at these directions, typically specified as a uniform gain corresponding to each direction.
The optimal beamforming vector $\boldsymbol{{w_{LS}}} \in \mathbb{C}^{ M_x^b M_y^b \times 1}$ is then obtained by solving the LS problem $\boldsymbol{{w_{LS}}} = {{\boldsymbol{{A_q}}}^\dag }\boldsymbol{v}$, where ${ \boldsymbol{A_q} ^\dag }$ denotes the pseudo-inverse of $\boldsymbol{A_q}$. This approach allows the synthesized beam pattern to approximate the desired spatial response in the least-squares sense, thereby enabling effective separation of multipath components.

After designing the receive beamformer $\boldsymbol{w_{LS}}$ for each path, the $l$-th multipath component can be separated by applying the corresponding beamformer to the received signals. Thus,
\begin{equation}\label{multi}
\begin{array}{l}
\begin{aligned}
{\hat{h}_{n,m}^l}=& \boldsymbol{{\left( {w_{LS}} \right)^H}}\boldsymbol{{\hat {h}_{n,m}}}\\
=& {h_{n,m}^l} + z_{n,m,h}^{l} \\
=& \sum\limits_{k \in {\Theta ^l}} {{\beta _{k}}{e^{j2\pi m{T_s}{f_k^e} }}} \times  {e^{ - j2\pi n\Delta f {{\tau _k^e}} }}{\chi _{t,k}}{\chi _{r,l,k}} + z_{n,m,h}^{l},
 \end{aligned}
\end{array}
\end{equation}
where ${h_{n,m}^l}$ denotes the actual channel response of $l$-th AoA on the $n$-th subcarrier of the $m$-th packet, ${{\Theta ^l}}$ denotes the set of targets located in the $l$-th AoA, ${\chi _{r,l,k}}$ denotes the receive beamforming gain of the $l$-th AoA and $z_{n,m,h}^{l}$ denotes the corresponding Gaussian noise vector. After stacking all $M_s \times N$ channel estimates, where $M_s$ denotes the number of packets and $N$ denotes the number of subcarriers, the $l \in \left\{ {0,1,2, \cdots ,{L}} \right\}$-th channel response matrix ${\boldsymbol{{\hat{H}^l}}} \in \mathbb{C}^{M_s\times N}$ is written as
\begin{equation}
 {\boldsymbol{{\hat{H}^l}}}  = \left[ {\begin{array}{*{20}{c}}
{\hat{h}_{0,0}^l}&{{\hat{h}_{1,0}^l}}&{ \cdots }&{{\hat{h}_{N-1,0}^l}}\\
{{\hat{h}_{0,1}^l}}&{{\hat{h}_{1,1}^l}}&{\cdots}&{{\hat{h}_{N-1,1}^l}}\\
{ \vdots }&{\vdots}&{ \ddots }&{\vdots}\\
{{\hat{h}_{0,M_s-1}^l}}&{{\hat{h}_{1,M_s-1}^l}}&{\cdots}&{{\hat{h}_{N-1,M_s-1}^l}}
\end{array}} \right]
\end{equation}

\begin{table}[!t]
	\centering
	\caption{Simulation Parameters \cite{int14,SM3,para1,MU1}}\label{simu}
	\begin{tabular}{c|c}
		\hline
		\hline
		\label{Parameter:simulation}
		{\textbf{Parameter}}  & {\textbf{Value}} \\
		\hline
		Total transmit power (${P_{total}}$)  & 30 dBm \\
        Carrier frequency (${f_c}$) & 28 GHz\\
        Carrier wavelength ($\lambda $) & ${c \mathord{\left/
 {\vphantom {c {{f_c}}}} \right.
 \kern-\nulldelimiterspace} {{f_c}}}$ \\
        Bandwidth ($B$)  & 100 MHz \\
        Subcarrier number ($N$) & 1024 \\
        Subcarrier interval ($\Delta f$) & ${B \mathord{\left/
 {\vphantom {B N}} \right.
 \kern-\nulldelimiterspace} N}$ \\
        Cyclic Prefix ($T_g$) & $3.34{e^{ - 6}}$ \\
        NLoS path number ($L$)  & 3 \\
		Speed of light ($c$) & $3 \times {10^8}$ m/s \\
		\hline
		\hline
	\end{tabular}
\end{table}

To construct the range-velocity map of targets in a bistatic ISAC system, a two-dimensional Fourier-based processing is performed on the estimated channel matrix. Specifically, an Inverse DFT (IDFT) is applied along each row of the matrix, which corresponds to the frequency domain samples across subcarriers within each OFDM packet. This transformation resolves the range profiles of the targets by exploiting the frequency-dependent phase shifts introduced by the target distances. Subsequently, a DFT is applied along each column, which spans multiple OFDM packets. Since the Doppler effect induces phase variations across time (i.e., across OFDM packets), this transformation enables the extraction of target velocity information through Doppler frequency estimation. The range-velocity map $\boldsymbol{S}$ is obtained as
\begin{equation}
\boldsymbol{S} = \boldsymbol{\rm {F_{col}}}  \cdot   {\boldsymbol{{\hat{H}^l}}}  \cdot  \boldsymbol{\rm {F_{row}^{-1}}}  ,
\end{equation}
where $\boldsymbol{\rm {F_{col}}} \in \mathbb{C}^{M_s\times M_s}$ is the DFT matrix applied to each column and $\boldsymbol{\rm {F_{row}^{-1}}} \in \mathbb{C}^{N \times N}$ is the IDFT matrix applied to each row.
The resulting 2D map $\boldsymbol{S}$ presents peaks at coordinates corresponding to the delay (range) and Doppler (velocity) of each target, thus enabling joint range-velocity estimation.
\section{Simulation Results}
In this section, we present numerical results to validate the performance of our proposed ISR-SE method in terms of noise estimation accuracy, training loss, and target detection accuracy. For this evaluation, we assume that the detecting UAV functions as the ISAC transmitter, transmitting signals for both communication and sensing while being connected to a ground BS for bistatic sensing tasks. The detecting UAV is capable of moving freely within an area of 1000 m $\times$ 1000 m based on the control commands from the ground BS, allowing it to optimize its trajectory for improved sensing performance. The ground BS, acting as the receiver, passively processes the reflected signals to extract the information of the targets, including angle, range, and velocity. The locations of the BS and the detecting UAV are set to (0, 0, 0) and (100, 100, 100), respectively. The three intrusion UAVs are randomly distributed within the bistatic sensing range. Additionally, the AWGN power at each receiver is set to ${\sigma _n ^2} = -174$ dBm/Hz, assuming a total transmit power of $P_{\rm{total}} = 1$ W. The detailed simulation parameters are shown in Table~\ref{simu}.

For the forward and reverse diffusion process, $T$ is set to 500. Too many steps may introduce a significant computational burden and memory consumption. The simulations revealed that fewer steps inadequately captured the noise characteristics, while significantly more steps excessively increased the computational complexity. Therefore, $T=500$ achieves an optimal balance between performance enhancement and computational burden. During training, the improved UNet denoising network is optimized using the Adam optimizer with a fixed learning rate of $10^{-4}$. The network is trained to minimize the loss, allowing it to accurately recover clean images from noisy inputs. Training is conducted over a predefined number of epochs with a constant learning rate to ensure a stable convergence. The training dataset was constructed from signal samples collected through our own measurement and acquisition process. This custom dataset is tailored to the characteristics of the bistatic ISAC system and includes a variety of signal scenarios under different SNR conditions, ensuring that the model is trained on representative and diverse data distributions.

\begin{figure}[!t]
	\centering
	\includegraphics[width=0.35\textheight]{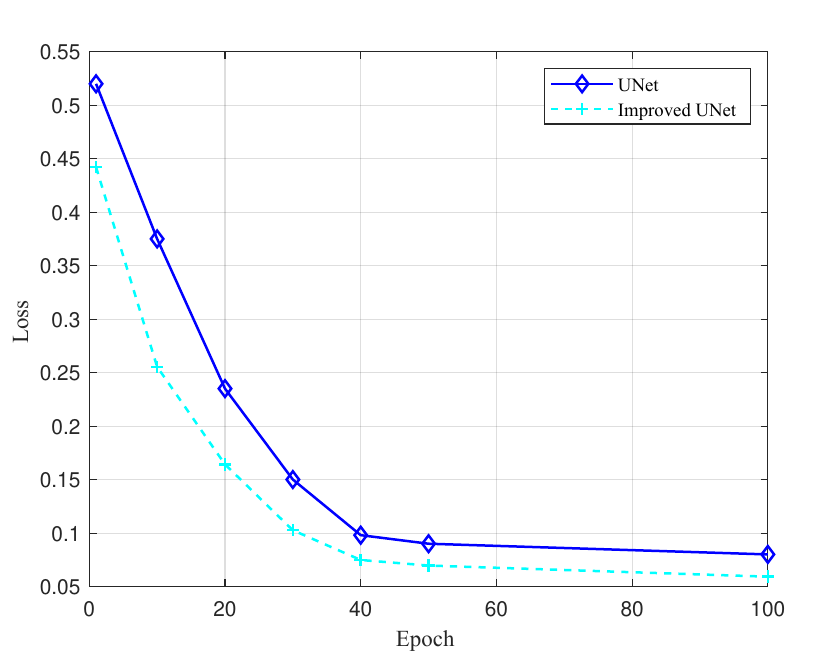}
	\DeclareGraphicsExtensions.
	\caption{The loss vs. training epochs under UNet and improved UNet denoising network.}
	\label{Imp}
\end{figure}

We first perform training optimization by exploring the impact of different training parameters on the performance of the improved UNet denoising network. Fig.~\ref{Imp} presents the loss comparison between the UNet denoising network and the proposed improved UNet architecture mentioned in Section~III, which incorporates residual connections and attention mechanisms into the UNet backbone. The results clearly show that the proposed model consistently achieves a lower loss across all training epochs. This improvement demonstrates the effectiveness of residual learning in preserving fine-grained details, and the attention mechanism in focusing on structurally important regions during denoising. In conclusion, the integration of residual and attention modules enables better representation learning and improved denoising performance in diffusion models.

\begin{figure}[!t]
	\centering
	\includegraphics[width=0.35\textheight]{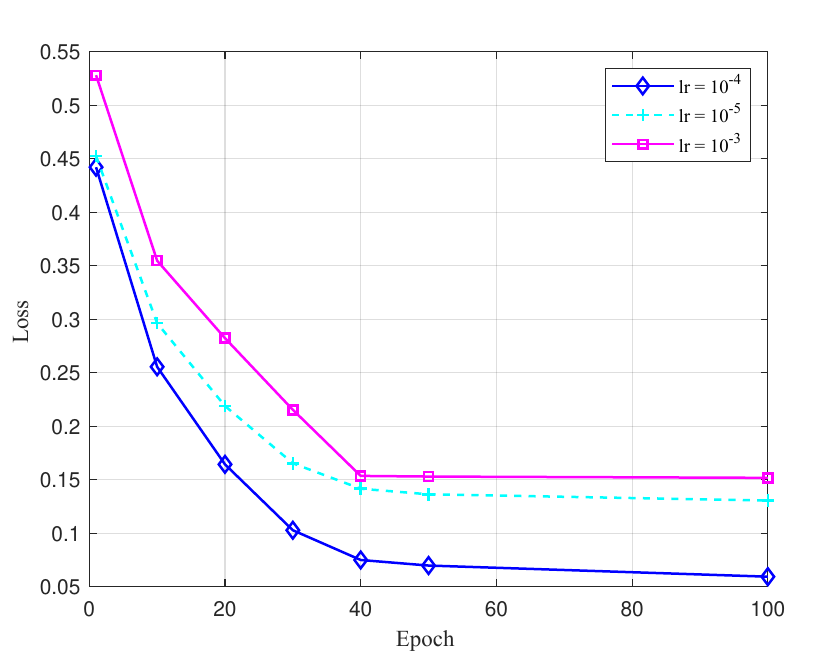}
	\DeclareGraphicsExtensions.
	\caption{The loss of proposed UNet denoising network vs. training epochs under different learning rates.}
	\label{LR}
\end{figure}

Fig.~\ref{LR} illustrates the variation of the training loss with respect to training epochs under different learning rates in the training of the proposed UNet denoising network. It is observed that a medium learning rate (lr = $10^{-4}$) achieves the best overall performance. It converges steadily and reaches the lowest loss, indicating an optimal balance between convergence speed and final accuracy. This learning rate enables the model to effectively minimize both the pixel-wise error and the structural distortion. A low learning rate (lr = $10^{-5}$) shows the most stable but slowest convergence. The loss decreases gradually, reflecting underfitting in the early stages. While this learning rate may be useful for fine-tuning, it is suboptimal for training from scratch due to a limited learning capacity per step. A high learning rate (lr = $10^{-3}$) provides a faster initial loss reduction but eventually plateaus at a higher final loss. This indicates that a large learning rate may hinder the model from reaching a better optimum, possibly due to oscillations or overshooting during gradient updates. In summary, the choice of learning rate plays a critical role in training stability and final model performance. Among the tested values, the medium learning rate offers the best trade-off and is recommended for most diffusion model training tasks.

Through systematic tuning, we identify the optimal configuration that yields the best denoising results. After obtaining the trained model, we conduct comparative experiments with the following methods to demonstrate the superior performance of our proposed method.

1) Traditional Signal Processing (TSP): Direct signal processing using a 2D-FFT and the MUSIC algorithm to estimate the information of the targets without any signal enhancement techniques.

2) Least Mean Square (LMS) \cite{LMS1,LMS2}: The adaptive filtering method based on the LMS error criterion can suppress noise interference and enhance the signal components by dynamically adjusting the filter coefficients.

3) Convolutional Neural Network (CNN) \cite{CNN1,CNN2}: By designing an appropriate network structure, a CNN can automatically learn and extract useful features from the input signal, so as to achieve signal enhancement.

To evaluate the effectiveness of our proposed ISR-SE method, we conduct a comparative analysis against the three aforementioned methods. The evaluation focuses on a key performance metric-Root Mean Squared Error (RMSE)-to quantify the accuracy of the estimation results. Specifically, the estimation Mean Squared Error (MSE) is computed as the average of the squared differences between the estimated and ground-truth values under a given simulation setting. The RMSE is then derived as the square root of the MSE, providing a more interpretable measure of estimation accuracy in the original signal scale.

\begin{figure}[!t]
	\centering
	\includegraphics[width=0.35\textheight]{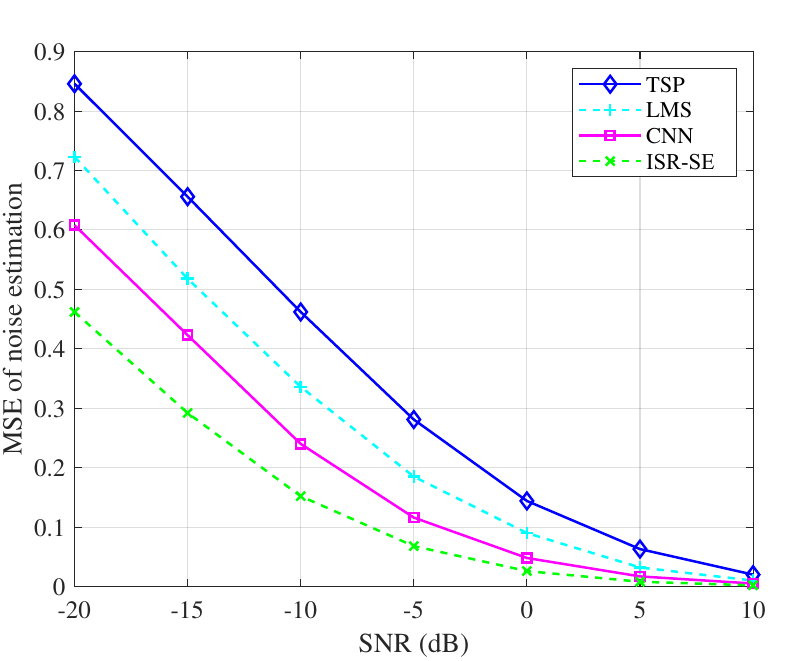}
	\DeclareGraphicsExtensions.
	\caption{The MSE of noise estimation vs. SNR using different signal enhancement methods.}
	\label{msenoise}
\end{figure}

First, we compare our proposed ISR-SE method with the three baseline methods in terms of the noise estimation loss. To quantify the performance, we adopt the MSE between the estimated and the ground-truth noise as the evaluation metric, which effectively reflects each method's noise suppression capability. The results, illustrated in Fig.~\ref{msenoise}, present the estimation loss across a wide range of SNR conditions. All four methods show a decreasing trend in MSE as the SNR increases. However, the rate and consistency of improvement vary according to the underlying algorithmic design. The TSP-based method, which lacks active noise mitigation mechanisms, exhibits the highest estimation loss, with MSE values up to three times larger than those of the other methods under low-SNR conditions, demonstrating its high sensitivity to noise. The LMS-based method delivers a stable performance in mid-SNR scenarios by adaptively updating the filter coefficients to minimize the noise. Nevertheless, due to its dependency on iterative convergence, it suffers from occasional error spikes in extreme low-SNR regimes, where its accuracy approaches that of the TSP-based approach.
The CNN-based method outperforms both TSP and LMS by leveraging data-driven learning to capture complex noise characteristics, thus achieving a lower estimation loss. However, its performance significantly deteriorates under severe noise levels, indicating limitations in its generalization ability. In contrast, the proposed ISR-SE method consistently yields the lowest estimation loss across all tested SNR levels. Notably, even under challenging conditions such as -20 dB, it maintains a superior accuracy, substantially outperforming all baseline methods. The most pronounced performance gains occur in the low-to-mid SNR range, where the ISR-SE method effectively enhances the signal components while suppressing the noise, highlighting its robustness and adaptability in complex environments.

\begin{figure}[!t]
	\centering
	\includegraphics[width=0.35\textheight]{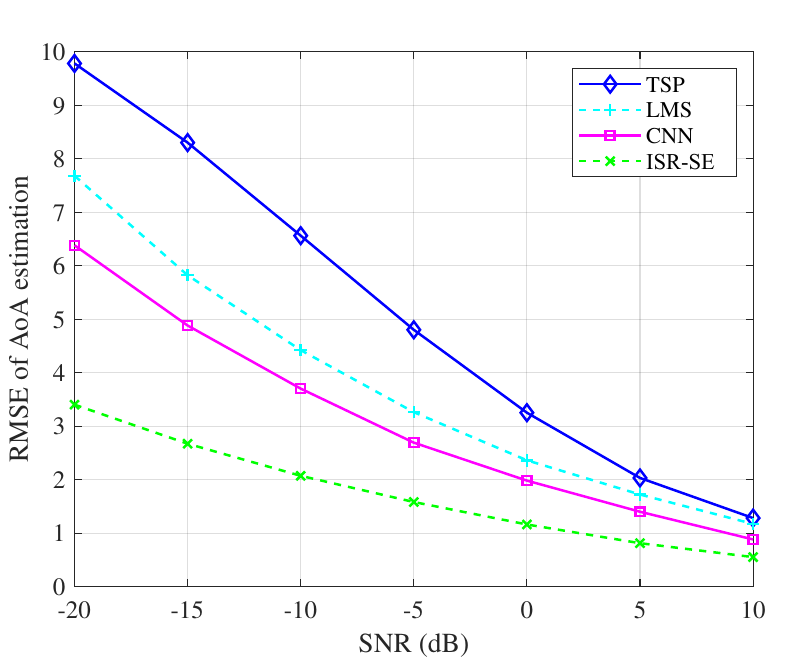}
	\DeclareGraphicsExtensions.
	\caption{The RMSE of AoA estimation vs. SNR using different signal enhancement methods.}
	\label{RMSEAoA1}
\end{figure}

\begin{figure}[!t]
	\centering
	\includegraphics[width=0.35\textheight]{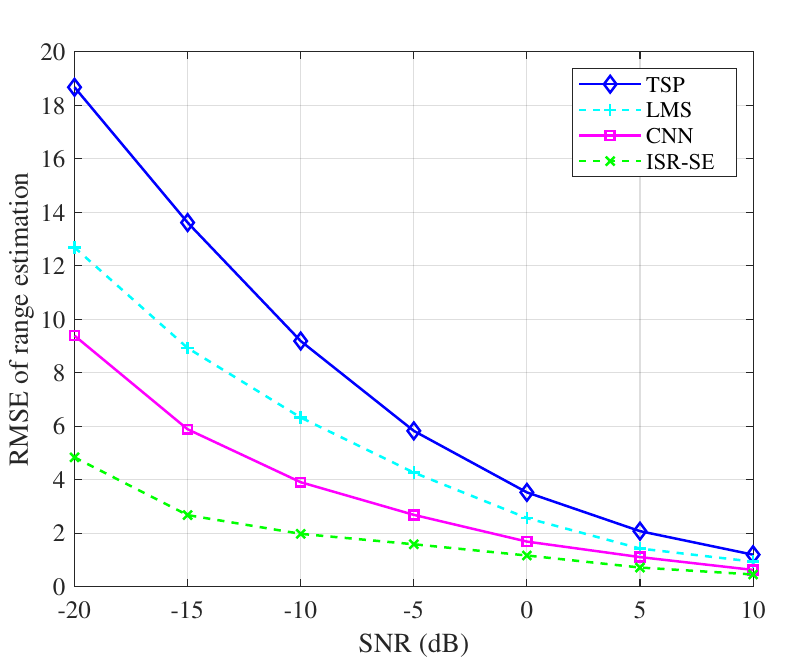}
	\DeclareGraphicsExtensions.
	\caption{The RMSE of range estimation vs. SNR using different signal enhancement methods.}
	\label{RMSERan1}
\end{figure}

\begin{figure}[!t]
	\centering
	\includegraphics[width=0.35\textheight]{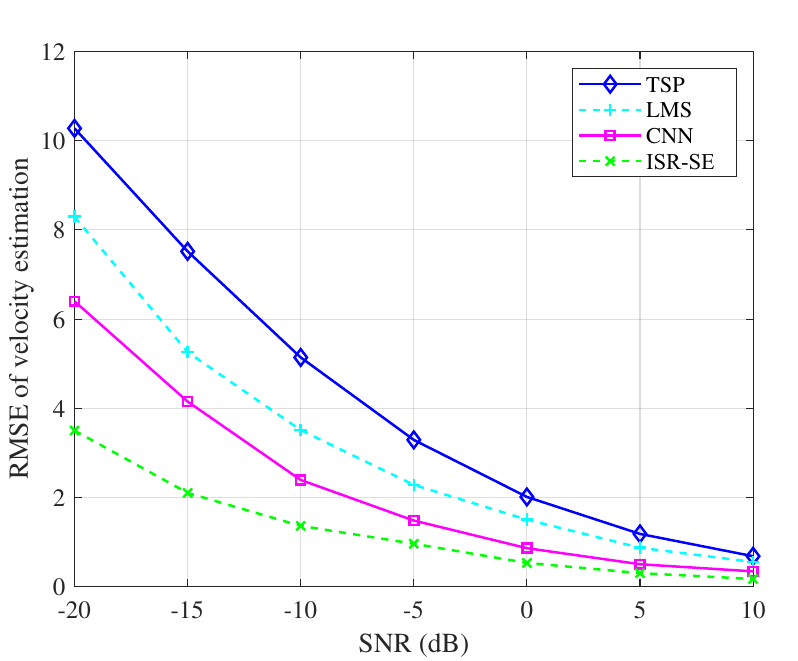}
	\DeclareGraphicsExtensions.
	\caption{The RMSE of velocity estimation vs. SNR using different signal enhancement methods.}
	\label{RMSEVel1}
\end{figure}

Fig.~\ref{RMSEAoA1}, Fig.~\ref{RMSERan1}, and Fig.~\ref{RMSEVel1} present the RMSE performance of AoA, range and velocity estimation under different signal enhancement methods across varying SNR levels. As expected, all methods exhibit a decreasing RMSE trend with increasing SNR, reflecting improved estimation accuracy under better signal quality. However, the rate of improvement and overall accuracy vary significantly across the different approaches. The TSP-based method consistently yields the highest RMSE for all three parameters (AoA, range, and velocity) across the entire SNR range, underscoring its sensitivity to noise. Under low-SNR conditions, the estimation RMSE remains markedly high, indicating a severe degradation in estimation accuracy. While performance improves at higher SNRs, it still lags behind the more advanced methods. The LMS-based method shows a moderate performance. It outperforms the TSP-based approach in mid-to-high SNR scenarios due to its adaptive filtering mechanism. However, under extremely noisy conditions, its estimation accuracy degrades significantly, revealing limitations in robustness and convergence efficiency in low-SNR environments. The CNN-based method achieves an improved performance compared to both TSP and LMS, particularly in high-SNR regimes. Its ability to model complex, non-linear signal patterns allows for enhanced estimation accuracy. Nevertheless, its effectiveness diminishes at extremely low SNRs, indicating limitations in generalization when facing severe noise. In contrast, the proposed ISR-SE method consistently outperforms all baselines, achieving the lowest estimation RMSE across the full range of SNRs. By leveraging an image-based denoising framework, the proposed ISR-SE method effectively preserves key signal structures while suppressing random noise, leading to substantial improvements in estimation accuracy. Notably, even at low SNRs (e.g., -20 dB), the proposed ISR-SE method maintains a significantly lower RMSE than the other methods. In particular, it reduces the estimation RMSE by over 63\% compared to the TSP-based method, highlighting its superior resilience to noise and robustness in adverse conditions.

\begin{figure}[!t]
	\centering
	\includegraphics[width=0.35\textheight]{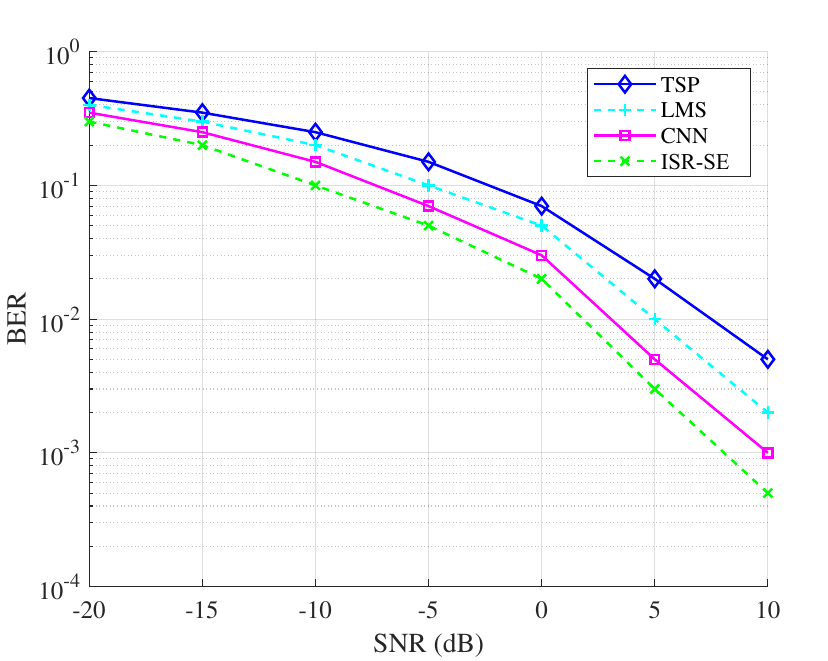}
	\DeclareGraphicsExtensions.
	\caption{The bit error rate of communication vs. SNR using different signal enhancement methods.}
	\label{BER1}
\end{figure}

Fig.~\ref{BER1} compares the Bit Error Rate (BER) performance of the four methods. The TSP-based method lacks active noise suppression mechanisms, making it highly vulnerable to channel impairments. Its direct signal processing without advanced filtering fails to mitigate noise interference effectively, which achieves the worst BER across all SNRs. While adaptive filtering improves noise suppression iteratively, the LMS-based method's convergence is severely hindered in extremely low-SNR environments, leading to an unstable performance. But it outperforms the TSP-based method in moderate-SNR regimes, reducing the BER from $10^{-1}$ to $10^{-3}$. The CNN-based method achieves near-optimal performance (BER$ \approx {10^{ - 4}}$) at high-SNR but degrades significantly at low-SNRs, lagging behind the proposed method by 1-2 orders of magnitude. The CNN-based models excel in learning noise patterns under trained conditions but struggle to generalize in extreme noise scenarios due to limited training diversity or insufficient robustness in feature extraction. The propsed ISR-SE method's innovative architecture addresses the limitations of both traditional adaptive algorithms and data-driven methods. By combining domain-specific denoising strategies and a multi-scale analysis, it achieves unmatched robustness across the entire SNR spectrum, making it ideal for low-SNR applications.

\begin{figure}[!t]
	\centering
	\includegraphics[width=0.35\textheight]{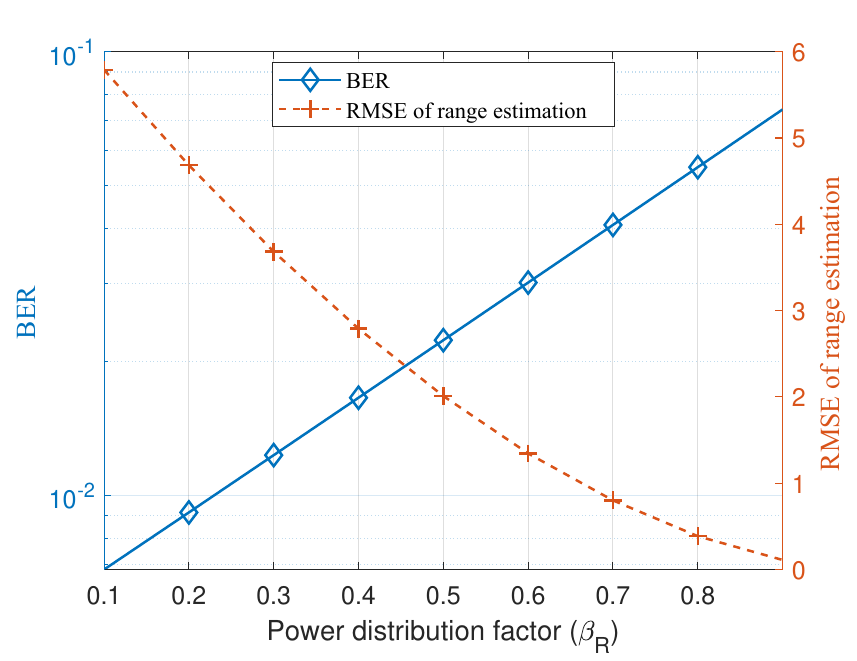}
	\DeclareGraphicsExtensions.
	\caption{The tradeoff of BER-range estimation performance vs. power distribution factor $\beta_R$.}
	\label{trade}
\end{figure}

To further illustrate the tradeoff between communication and sensing performance influenced by the power allocation factor $\beta_R$, we performed additional simulations focusing on realistic operational scenarios. Specifically, we varied $\beta_R$ within a practical range, avoiding the extreme cases of full allocation to either communication or sensing. Fig.~\ref{trade} clearly demonstrates the inherent tradeoff. As $\beta_R$ increases, allocating more power toward sensing tasks, the BER gradually deteriorates to the $10^{-1}$ level. This exponential increase in BER indicates a notable degradation in communication reliability due to reduced transmit power for communication signals. Conversely, the range estimation accuracy (measured by the RMSE) significantly improves with an increasing sensing power allocation. The RMSE of the range decreases substantially as expected. However, it is also clear that the improvement exhibits diminishing returns, with the most pronounced accuracy enhancement observed at moderate allocation levels. This comprehensive analysis underscores the necessity of carefully selecting the power allocation factor to balance the conflicting demands of high-quality communication and accurate sensing in practical ISAC systems.

\section{Conclusion}
In this paper, we consider a bistatic ISAC system consisting of a ground BS and a detecting UAV collaboratively performing bistatic sensing tasks. We propose a novel image super-resolution-based signal enhancement framework specially designed for bistatic ISAC systems operating in low-SNR environments. By transforming the received signals into RGB representations via a STFT-based time-frequency analysis, and subsequently enhancing these images through a hybrid denoising network that combines the strengths of improved UNet and diffusion model architectures, our proposed ISR-SE method significantly improves the signal clarity and the robustness against noise. Leveraging the powerful generative capabilities of GAI, the proposed ISR-SE method enables accurate extraction of communication symbols and high-precision estimation of target parameters such as range, velocity, and angle. Extensive simulation results validate the effectiveness of the proposed ISR-SE method, demonstrating a notable improvement of up to 63\% in estimation accuracy compared to traditional approaches. By leveraging generative models, AI-driven frameworks are capable of learning intricate patterns hidden in raw or transformed signal domains, thus offering superior resilience under low-SNR, interference-rich, or incomplete-data conditions. Therefore, the integration of AI with wireless signal processing is not only a technical enhancement but a foundational enabler for the realization of the fully intelligent, autonomous, and integrated wireless systems envisioned in the 6G era.


\ifCLASSOPTIONcaptionsoff
  \newpage
\fi

\end{document}